%% file: main.tex
\documentclass[journal]{IEEEtran}

\pdfoutput=1

\usepackage{cite}
\usepackage{url}

\ifCLASSINFOpdf
    \usepackage[pdftex]{graphicx}
\else
    \usepackage[dvips]{graphicx}
\fi

\usepackage{amsmath}
\usepackage{amssymb}
\input{global_math_definitions}

\input{units}

\begin{document}

\title{Millimeter-Wave Transmission-Line Reflectionless Filters}
%
%
%

\author{Matthew~A.~Morgan,~\IEEEmembership{Senior Member,~IEEE,}
        Matt Bauwens,
        Seng Loo,
        Miho Hunter,
        Tod~A.~Boyd,
        and Robert M. Weikle,~\IEEEmembership{Fellow,~IEEE}
\thanks{Manuscript received 11/11/2025}
\thanks{M. Morgan and T. Boyd are with the Central Development Laboratory, National Radio Astronomy Observatory, Charlottesville,
VA, 22903 USA (e-mail: matt.morgan@nrao.edu). The National Radio Astronomy Observatory is a facility of the National Science Foundation operated under cooperative agreement by Associated Universities, Inc.}
\thanks{Matt Bauwens and Robert M. Weikle are with Dominion Microprobes Inc. (DMPI), Charlottesville, VA, 22905. Robert M. Weikle is also associated with the Unversity of Virginia, Charlottesville, VA, 22904.}
\thanks{Seng Loo and Miho Hunter are with Anritsu Company, Morgan Hill, CA, 95037 USA.}}

\markboth{IEEE Microwave and Wireless Components Letters,~Vol.~xx, No.~x, xxx~2026}%
{Morgan \MakeLowercase{\textit{et al.}}: Millimeter-Wave transmission-Line Reflectionless Filters}

\IEEEpubid{0000--0000/00\$00.00~\copyright~2018 IEEE}


\maketitle

\begin{abstract}
We report on the development of transmission-line reflectionless filters operating with passbands at 100 GHz and 230 GHz, and stopband absorption up to 500 GHz, the highest operating frequencies yet recorded for such filters. The designs are based on a previously reported mathematical solution to the reflectionless condition, now successfully implemented for the first time, using an advanced thin-film fabrication process on Alumina substrates. Sub-millimeter wave wafer probe measurements show good agreement with theory.\end{abstract}


\section{Introduction}

\IEEEPARstart{D}{espite} some initial work in the early part of the twentieth century on constant-resistance networks for telephony \cite{zobel1928,bode}, the subject of absorptive filtering had, by the time of the wireless revolution, fallen into relative obscurity, with only occasional publications on empirical approaches appearing in the academic literature prior to 2011. Publication of the coupled-ladder topology in that year \cite{morgan_theoretical}, however, triggered a resurgence of interest that has grown globally into an active area of research. Dozens of papers are now published every month on new reflectionless and absorptive filter topologies \cite{gomez-garcia2020dec}, synthesis methods \cite{khalajamirhosseini2016,lee2020jun_tmtt,guilabert2019}, implementations \cite{xu2023,he2023,xiao2024,ma2017}, and applications \cite{kow2025,kedziora2025,li2024,thanh2023_english}, as researchers look to reflectionless filters as a solution to the problems associated with out-of-band standing waves, reflected noise, spurious mixing products, harmonic generation, and image-band termination.


\section{Transmission-Line Topology}

\begin{figure}[tbp]
    \centering
    \includegraphics{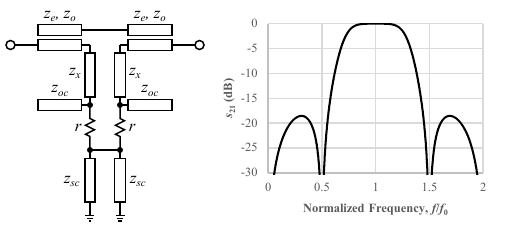}\\
    (a)\hspace{1.6in}(b)\hspace{0.2in}
    \caption{(a)~Transmission-Line reflectionless filter topology. Lower-case parameters indicate normalized impedance values. (b)~Simulated performance with ideal elements and $z_x=\sqrt{2}$. Return loss in dB is infinite.}
    \label{fig:topology}
\end{figure}

The transmission-line topology we have used is shown in Fig.~\ref{fig:topology}(a).
Like the lumped-element coupled-ladder topologies already published \cite{morgan_theoretical,morgan_ladder,morgan_thin-film_reflectionless}, it was proposed as a solution to the reflectionless condition,
\begin{equation}
    s_{ii}(f) = 0\quad\forall i,f
\end{equation}
That is, the schematic model, given ideal elements without parasitics, has identically zero reflection coefficient at all frequencies and from all ports. It was shown \cite{morgan_structures,morgan_artech,morgan9923540,morgan10277189,morgan6964152B2,morgan6652970B2} that this criterion is satisfied when the lines are equal in phase and have the following characteristic impedances,
\begin{myeqn}
    z_{sc} &=& z_x-z_x^{-1}\\
    z_{oc} &=& z_{sc}^{-1}\\
    \rho &=& 1+2z_{sc}+2\sqrt{z_{sc}\left(1+z_{sc}\right)}\\
    z_e &=& 2\rho/\left(\rho+1\right)\\
    z_o &=& 2/\left(\rho+1\right)
\end{myeqn}
where the lower-case $z$'s indicate normalized impedance values. The free parameter $z_x$, then, determines the bandwidth centered around the frequency at which the lines are a quarter-wavelength long. A plot of the theoretical $s$-parameters is shown in Fig.~\ref{fig:topology}(b) for $z_x=\sqrt{2}$.

\IEEEpubidadjcol

Although theoretically sound, implementation of this topology since its initial publication has proven difficult in practice due primarily to the rather extreme impedances and tight coupling required. With $z_x=\sqrt{2}$ and $Z_0=50\Omega$, we have
\begin{myeqn}
    Z_x &\approx& 70.7\Omega\\
    Z_{sc} &\approx& 35.4\Omega\\
    Z_{oc} &\approx& 70.7\Omega\\
    Z_e &\approx& 82.2\Omega\\
    Z_o &\approx& 17.8\Omega
\end{myeqn}
where the capital $Z$'s now indicate actual, rather than normalized, impedance values. Increasing $z_x$ tightens the coupling of the already challenging coupled lines, while reducing it pushes the open and short-circuited stub impedances further away from easily realizable values.

Additionally, the prior success of very compact lumped-element reflectionless filters up to $60\GHz$ \cite{morgan_thin-film_reflectionless} ensures that this alternative transmission-line topology will have its greatest impact enabling operation at much higher frequencies. The designer must therefore be prepared to implement this filter on electrically thick substrates.

\section{Layout and Fabrication}

\begin{figure}[tbp]
    \centering
    \includegraphics{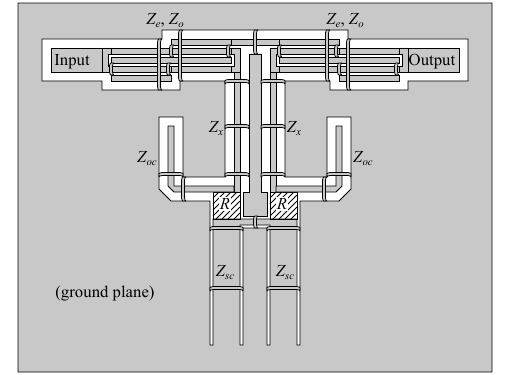}
    \caption{Layout plan for transmission-line reflectionless filters. Lange hybrids are used to implement the coupled line sections, while CPW is used for the remainder of the transmission lines and stubs.}
    \label{fig:layout}
\end{figure}

To address these challenges, we chose to implement our prototypes using an advanced thin-film process capable of achieving tight tolerances with small features, to implement the coupled lines using a Lange/interdigital coupling section \cite{lange1969}, and to realize the high and low-impedance transmission lines on the top surface of the substrate using coplanar waveguide (CPW). The general layout is thus shown in Fig.~\ref{fig:layout}.
Air bridges are used at each bend and no less than $\lambda/8$ apart, including over the interdigital sections of the Lange couplers, to ensure that all areas of the ground plane remain at the same electric potential. Alumina (Al$_2$O$_3$) was chosen as the substrate material, as it has a relatively high dielectric constant ($\epsilon_r=9.9$), making it easier to achieve the tight coupling needed in the coupled-line sections.

\begin{figure}[tbp]
    \centering
    \includegraphics{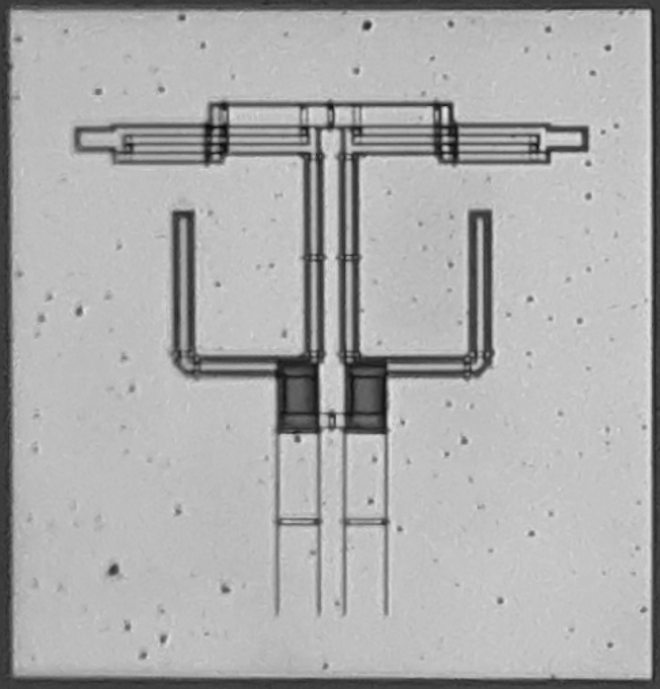}\hfill\raisebox{0.5in}{\includegraphics{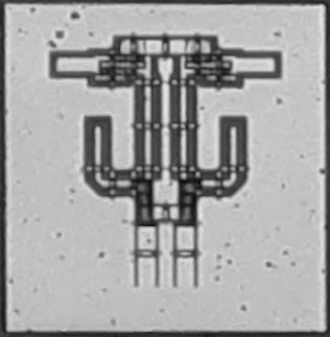}}\\
    \hspace{0.6in}(a)\hspace{1.7in}(b)
    \caption{Microphotographs of transmission-line reflectionless filters at true relative scale. (a)~$100\GHz[-]$ bandpass filter. Chip dimensions are $850\um\times900\um$. (b)~$230\GHz[-]$ bandpass filter. Chip dimensions are $450\um\times450\um$.}
    \label{fig:photos}
\end{figure}

Two designs were completed, one with a passband center frequency of $100\GHz$, and another centered at $230\GHz$. Microphotographs are shown in Fig.~\ref{fig:photos}.
They were fabricated together on a $150\mm[-]$ Alumina wafer, with a metal stackup comprising two metal layers, $1.25\um$ and $2.5\um$ thick, respectively, with $4\um[-]$ thick polyimide bridges, and resistors formed from TaN film having a sheet resistance of $50\Omega/\square$. The wafer yielded approximately $36,000$ die, or $18,000$ of each design.

\section{Measurements}

Wafer probe measurements were taken from sites all over the wafer---one of each design from alternating reticules, or a total of nearly 50 die each. A small number of outliers from the edges of the wafer were excluded.

The $100\GHz[-]$ design was tested from $1-220\GHz$ using FormFactor T220-UBBT-GSG-50, $50\um[-]$ pitch dual-band T-Wave probes on a FormFactor PA200 semi-automatic probe station, and a Keysight PNA-X with $1\mm[-]$ coax and WR-5.1 waveguide frequency extenders. The $230\GHz[-]$ design required testing in three bands. The first measurement band was from $1-220\GHz$ using the same setup described above. The second measurement band extended from $220-325\GHz$, and was performed again with a Keysight PNA-X, this time with WR-3.4 frequency extenders, model WR3.4VNATxRx from Virginia Diodes (VDI), and FormFactor T330-S-GSG-25-BT, $25\um[-]$ pitch T-Wave probes. Finally, band 3 covered from $325-500\GHz$ using VDI WR2.2VNATxRx WR-2.2 frequency extenders and FormFactor T500-S-GSG-25-BT, $25\um[-]$ pitch WR-2.2 T-Wave probes. In all cases, TRL calibration was performed using FormFactor calibration substrates, placing the reference planes at the probe tips.

\begin{figure}[tbp]
    \centering
    \includegraphics{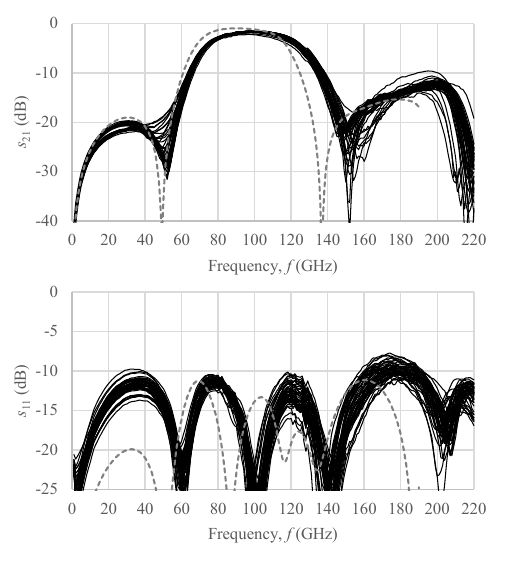}
    \caption{Plots of $s_{21}$ and $s_{11}$ for $100\GHz[-]$ reflectionless filters. Measured curves are shown with solid black lines, while the EM-simulated results are shown with dashed gray lines.}
    \label{fig:lowfreq_plot}
\end{figure}

\begin{figure}[tbp]
    \centering
    \includegraphics{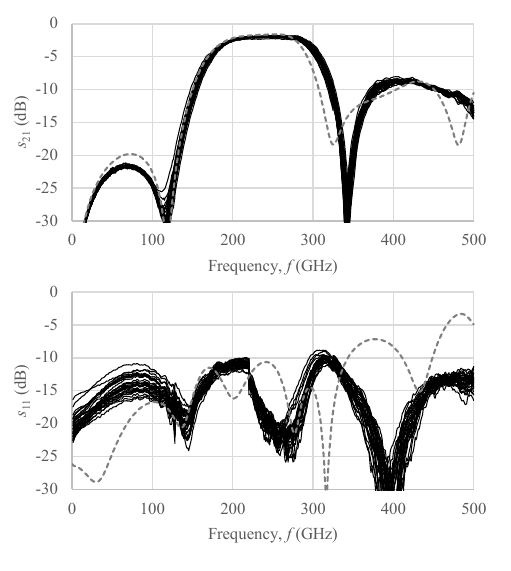}
    \caption{Plots of $s_{21}$ and $s_{11}$ for $230\GHz[-]$ reflectionless filter. Measured curves are shown with solid black lines, while the EM-simulated results are shown with dashed gray lines.}
    \label{fig:hifreq_plot}
\end{figure}

The final data are shown for the $100\GHz[-]$ design in Fig.~\ref{fig:lowfreq_plot},
and for the $230\GHz[-]$ design in Fig.~\ref{fig:hifreq_plot}.
Measured data are plotted with solid black lines and electromagnetic (EM) simulated data with dashed gray lines. Note that the simulations performed using Keysight's Advanced Design System (ADS) confirmed the $20\dB[-]$ attenuation in the lower-stopband from the ideal element model in Fig.~\ref{fig:topology}, while accurately predicting a degradation in the upper stopband attenuation, from $20\dB$ to $15\dB$ in the case of the lower frequency design, and from $20\dB$ to $10\dB$ for the higher frequency design. This was a consequence of the transmission-line dispersion properties on a high-dielectric constant substrate.

Uniformity among die across the wafer is excellent, and agreement between simulation and measurement is reasonably good in both cases, with some noted differences. In particular, both die exhibit an upward shift in passband frequency, especially at the upper transition edge (about $10\%$ and $6\%$ for the lower and higher-frequency designs, respectively). This may be attributed to the polyimide layer, which was only present beneath the bridges in the actual fabricated chips, but lay across the entire surface of the design in simulation due to limitations of the EM software license available. This may also account for the offset locations of the reflection zeros in the return loss curves. Finally, we note a small discontinuity in the return loss data for the high-frequency design between the first and second measurement bands at $220\GHz$, indicating imperfection in the probe calibration at the band edges.

\begin{figure}
    \centering
    \includegraphics{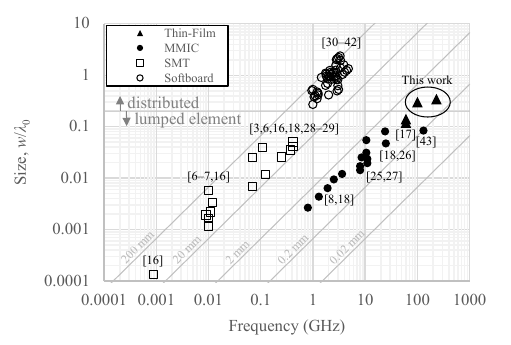}
    \caption{Comparison of reflectionless filter implementations using a variety of techniques and technologies, showing chip size in free-space wavelengths vs. operating frequency. The horizontal line at $0.2\lambda_0$ divides lumped-element designs from distributed/transmission-line topologies. (Transitions and bond/probe pads are included in ccircuit dimensions.)}
    \label{fig:comparison}
\end{figure}

These filters are compared against a broad sample of other absorptive and reflectionless filters \cite{morgan_thin-film_reflectionless, simpson2021sep,ge2021,xu2023,zhao2023,morgan_structures,morgan_theoretical,khalajamirhosseini2017,lee2020jun_tmtt,guilabert2019,psychogiou2019,morgan_ladder,wu2020aug,gomez-garcia2019sep,gomez-garcia2019apr,gomez-garcia2020apr,gomez-garcia2018nov,gomez-garcia2018apr,psychogiou2018jan, gomez-garcia2018sep,yang2020mar,psychogiou2020aug,zhao2022,lee2021dec,fan2021jan,sauber2024} in Fig.~\ref{fig:comparison}.
The plot shows the scale size in free-space wavelengths versus operating frequency at the center of the passband for bandpass filters, and at the cutoff corner for lowpass or highpass filters. Thin-film and MMIC technologies, shown with the filled triangles and circles, respectively, achieve the smallest physical dimensions and highest frequencies, while surface mount (SMT) solutions achieve the smallest relative scales according to wavelength.\footnote{Also noted is the curious lack of designs, of any kind, between $2\mm$ and $20\mm$ in size.} The filters described in this work occupy the far right of the plot, representing at once the highest frequency designs yet demonstrated and among the smallest per wavelength of any distributed approach.

\section{Conclusion}

Two millimeter-wave transmission-line reflectionless filters with passbands centered at $100\GHz$ and $230\GHz$ have been designed, fabricated on Alumina using a thin-film process, and tested up to $500\GHz$ by wafer probes. These are the highest frequency bandpass reflectionless filters ever reported. They are also among the smallest ever reported for any distributed element design, both physically and relative to wavelength.


\IEEEtriggeratref{25}

\bibliographystyle{morgan}
\bibliography{IEEEabrv,morgan_abrv,morgan,wiki,foreign,other} 


\end{document}

%% file: global_math_definitions.tex
\newenvironment{myeqn}[1][rCl]
    {\allowdisplaybreaks\begin{IEEEeqnarray}{#1}\IEEEyesnumber\IEEEyessubnumber*}
    {\end{IEEEeqnarray}\ignorespacesafterend}

%% file: units.tex
\newcommand{\GHz}[1][ ]{\text{{#1}GHz}}

\newcommand{\dB}[1][ ]{\text{{#1}dB}}

\newcommand{\mm}[1][ ]{\text{{#1}mm}}

\newcommand{\um}[1][ ]{\text{{#1}}\mu\text{m}}

